\documentstyle[12pt,subeqn]{article}

\newcommand\vek[1]{\mbox{\rmfamily\bfseries\itshape#1}}
\newcommand\vekexp[1]{\mbox{\scriptsize\rmfamily\bfseries\itshape#1}}

\begin{document}

\title{Coulomb Systems with Ideal Dielectric Boundaries:
Free Fermion Point and Universality}

\author{
B. Jancovici$^1$ and L. {\v S}amaj$^{1,2}$
}

\maketitle

\begin{abstract}
A two-component Coulomb gas confined by walls made of ideal
dielectric material is considered.
In two dimensions at the special inverse temperature 
$\beta = 2$, by using the Pfaffian method, the system is mapped 
onto a four-component Fermi field theory with specific
boundary conditions.
The exact solution is presented for a semi-infinite geometry
of the dielectric wall (the density profiles, the correlation
functions) and for the strip geometry (the surface tension,
a finite-size correction of the grand potential).
The universal finite-size correction of the grand potential 
is shown to be a consequence of the good screening properties,
and its generalization is derived for the conducting Coulomb
gas confined in a slab of arbitrary dimension $\ge 2$ at
any temperature.
\end{abstract}


\medskip
\noindent LPT Orsay 00-138

\vfill

\noindent $^1$ Laboratoire de Physique Th{\'e}orique, Universit{\'e}
de Paris-Sud, B{\^a}timent 210, 91405 Orsay Cedex, France (Unit{\'e}
Mixte de Recherche no. 8627 - CNRS);

\noindent e-mail: Bernard.Jancovici@th.u-psud.fr 
and Ladislav.Samaj@th.u-psud.fr

\noindent $^2$ On leave from the Institute of Physics, Slovak Academy
of Sciences, Bratislava, Slovakia

\newpage

\section{Introduction}
Very recently, the interface between a Coulomb gas and a dielectric
wall with zero dielectric constant (termed ideal dielectric) has been
studied by one of us (L.{\v S}.) \cite{Samaj1} in the case of an exactly
solvable two-dimensional model \cite{Samaj2}: 
a two-component plasma of point-particles. 
The surface tension has been obtained, for an arbitrary temperature 
within the range of surface stability of the model $\beta<3$
($\beta$ is the inverse temperature). 
This range exceeds the bulk range of stability $\beta<2$.
However, the method used in \cite{Samaj1}, a mapping onto a solvable 
boundary sine-Gordon field theory, provided only the surface tension 
but did not give any information about the microscopic structure of 
the interface: density profiles and correlation functions.

It can be expected that this microscopic information is obtainable at
the special inverse temperature $\beta=2$. 
This is because, at this special temperature, for solving other
problems, the model can be mapped onto a free fermion field theory
(This temperature is the boundary of the stability domain of the model: 
for point particles, at a given fugacity, the bulk density and 
other bulk thermodynamic quantities diverge, however they can be made 
finite by the introduction of a small hard core in the interaction. 
Anyhow the \emph{truncated} many-body densities are finite, even for 
point-particles). 
The two-dimensional two-component plasma at $\beta=2$ has been 
extensively studied, in the bulk \cite{Cornu1}, and also near 
a variety of possible interfaces, in particular the interface with 
a plain (simple) hard wall \cite{Cornu2} and with an ideal conductor 
wall \cite{Cornu2} -- \cite{Jancovici}. 
However, the case of the interface with an ideal dielectric could not 
be solved by the simplest possible extension of the mapping on 
free fermions with an inhomogeneous ``mass'' term 
which had been used for other interfaces. 

In the present paper, we do solve the problem of the interface with an
ideal dielectric by an appropriate generalization of the mapping. 
In the previously studied cases, the mapping was obtained by associating 
a two-component Fermi field with each point of space. 
In the present case, inspired by a remotely related reference 
\cite{Sheng}, we realized that a mapping on a four-component 
Fermi field is needed. 
Through that mapping, we are able to compute the density profile 
and the correlation functions near the interface. 
This is described in Section 2 dealing with a semi-infinite
geometry of the dielectric wall.

In Section 3, we turn to the problem of the two-dimensional model
at $\beta=2$ confined in a strip of width $W$ between two ideal 
dielectric walls. 
We show that the grand potential has the same universal finite-size 
correction of order $1/W$ as in the previously studied case of 
ideal conductor walls \cite{Jancovici}. 
Actually, as shown in Section 4, this finite-size correction is 
a consequence of the good screening properties and its generalization 
can be derived for a conducting Coulomb system of arbitrary dimension 
at any temperature.

\renewcommand{\theequation}{2.\arabic{equation}}
\setcounter{equation}{0}

\section{Model and method of solution}

\subsection{Pfaffian method}
We consider an infinite 2D space of points ${\vek r} \in R^2$
defined by Cartesian $(x,y)$ or complex $(z=x+{\rm i}y,
{\bar z}=x-{\rm i}y)$ coordinates.
The model dielectric-electrolyte interface is localized along
the $x$ axis at $y=0$.
The half-space $y<0$ is assumed to be occupied by an ideal 
dielectric wall of dielectric constant $\epsilon =0$, 
impenetrable to particles.
The electrolyte in the complementary half-space $y>0$
is modeled by the classical 2D two-component plasma
of point particles $\{ j\}$ of charge $\{ q_j = \pm 1\}$,
immersed in a homogeneous medium of dielectric constant $=1$
and interacting via Coulomb interaction.

At the special inverse temperature $\beta =2$, in order to avoid
the collapse of positive and negative particles, we start
with a lattice version of the Coulomb model.
There are two interwoven sublattices of sites $u\in U$
and $v\in V$.
The positive (negative) particles sit on the sublattice $U$
($V$); each site can be either empty or occupied by one particle.
We work in the grand canonical ensemble, with position-dependent
fugacities $\zeta(u)$ and $\zeta(v)$ in order to generate
multi-particle densities.
In infinite space, the Coulomb potential $v_0$ at spatial
position ${\vek r}$, induced by a charge at the origin,
is $v_0({\vek r}) = - \ln ( \vert{\vek r}\vert /a)$, where
the length constant $a$ fixes the zero point of energy.
The presence of a dielectric wall induces to a particle
at ${\vek r} = (x,y)$, or $z$, an image at ${\vek r}^* = (x,-y)$,
or ${\bar z}$, of the same charge \cite{Jackson}.
We consider only neutral configurations of $N$ positive and 
$N$ negative particles,
the complex coordinates of which are $u_i$ and $v_i$, respectively.
The corresponding Boltzmann interaction factor, 
involving direct particle-particle interactions 
$q_i q_j v_0(\vert {\vek r}_i - {\vek r}_j \vert)$
and interactions of particles with the images of other
particles $q_i q_j v_0(\vert {\vek r}_i - {\vek r}^*_j \vert)$
as well as their self-images 
$(1/2)q^2_i v_0(\vert {\vek r}_i - {\vek r}_i^* \vert)$, 
is, at $\beta =2$,
\begin{eqnarray} \label{2.1}
& & B_N(\{ u_i\};\{ v_i\}) =  (-1)^N a^{2N} \prod_i 
(u_i - {\bar u}_i) (v_i - {\bar v}_i)  \nonumber \\
& & \quad {\prod_{i<j} (u_i - u_j) 
({\bar u}_i - {\bar u}_j)(u_i - {\bar u}_j) ({\bar u}_i - u_j)
(v_i - v_j) ({\bar v}_i - {\bar v}_j)(v_i - {\bar v}_j) 
({\bar v}_i - v_j) \over
\prod_{i,j} (u_i - v_j) ({\bar u}_i - {\bar v}_j)
(u_i - {\bar v}_j) ({\bar u}_i - v_j)}
\end{eqnarray}
By using an identity of Cauchy \cite{Aitken}
\begin{subequations} \label{2.2}
\begin{equation} \label{2.2a}
{\rm det} \left( {1\over z_i - z'_j} \right)_{i,j=1,\ldots,2N}
= (-1)^{N(2N-1)} {\prod_{i<j}(z_i-z_j) (z'_i-z'_j) \over
\prod_{i,j} (z_i-z'_j)}
\end{equation}
with the identification
\begin{equation} \label{2.2b}
z_{2i-1} = u_i, \quad z_{2i} = {\bar u}_i, \quad z'_{2i-1} = v_i,
\quad z'_{2i} = {\bar v}_i, \quad \quad \quad i = 1,\ldots,N
\end{equation}
\end{subequations}
the Boltzmann factor (\ref{2.1}) is expressible as the determinant
of an $N\times N$ matrix whose elements are $2\times 2$ matrices:
\begin{equation} \label{2.3}
B_N(\{ u_i\};\{ v_i\}) =  
{\rm det} 
\left( \begin{array}{cc}
\displaystyle{a\over u_i-v_j} & 
\displaystyle{a\over u_i-{\bar v}_j} \\
\displaystyle{a\over {\bar u}_i-v_j} & 
\displaystyle{a\over {\bar u}_i-{\bar v}_j}
\end{array} \right)_{i,j=1,\ldots,N}
\end{equation}
The grand partition function reads
\begin{equation} \label{2.4}
\Xi = 1 + \sum_{u\in U \atop v \in V} \zeta(u) \zeta(v) B_1(u;v)
+ \sum_{u_1<u_2\in U \atop v_1<v_2\in V} \zeta(u_1) \zeta(u_2)
\zeta(v_1) \zeta(v_2) B_2(u_1,u_2;v_1,v_2) + \ldots
\end{equation}
where the sums are defined in such a way that configurations 
which differ by a permutation of identical-charge particles 
are counted only once.

We introduce a couple of Grassmann variables for every
site $u\in U$, $(\psi_u^1, \psi_u^2)$, and every site 
$v\in V$, $(\psi_v^1, \psi_v^2)$, and order all of them into
a single vector $^t\theta = ( \ldots \psi_u^1, \psi_u^2 \ldots
\psi_v^1, \psi_v^2 \ldots )$ whose components obey the ordinary
anticommutation rules $\{ \theta_i, \theta_j \} = 0$.
Let us consider the Grassmann integral
\begin{subequations} \label{2.5}
\begin{equation} \label{2.5a} 
Z_0 = \int {\rm d} \theta \ {\rm e}^{-S_0}
\end{equation}
where ${\rm d}\theta = \prod_v {\rm d} \psi_v^2 {\rm d} \psi_v^1 
\prod_u {\rm d} \psi_u^2 {\rm d} \psi_u^1$ and
\begin{equation} \label{2.5b}
-S_0 = {1\over 2}\ ^t\theta {\bf A} \theta
\end{equation}
\end{subequations}
is (minus) the Gaussian action with an antisymmetric matrix
${\bf A}$, $A_{ij} + A_{ji} = 0$.
We recall that ${\bf A}^{-1}$ is also antisymmetric.
The averaging with $S_0$ will be denoted by
\begin{equation} \label{2.6}
\langle \ldots \rangle = {1\over Z_0} \int {\rm d}\theta
\ \ldots {\rm e}^{-S_0}
\end{equation}
Gaussian integrals of type (\ref{2.5}) are expressible as
\cite{Zinn-Justin}
\begin{equation} \label{2.7}
Z_0 = {\rm Pf}({\bf A})
\end{equation}
where ${\rm Pf}({\bf A})$ is the Pfaffian of the antisymmetric
matrix ${\bf A}$, satisfying the well known identity
\begin{equation} \label{2.8}
{\rm Pf}({\bf A})^2 = {\rm det}{\bf A}
\end{equation}
The two-variable averages are given by
\begin{equation} \label{2.9}
\langle \theta_i \theta_j \rangle = (A^{-1})_{ji}
\end{equation}
The standard Wick theorem for fermions generalizes to
higher-order averages of $\theta$ variables \cite{Ramond}.
The ${\bf A}$-matrix will be chosen such that its inverse
satisfies the following relations:
\begin{subequations} \label{2.10}
\begin{eqnarray} 
(A^{-1})_{uu'}^{\alpha\alpha'} & = & \langle \psi_{u'}^{\alpha'}
\psi_u^{\alpha} \rangle = 0 \label{2.10a} \\
(A^{-1})_{vv'}^{\alpha\alpha'} & = & \langle \psi_{v'}^{\alpha'}
\psi_v^{\alpha} \rangle = 0 \label{2.10b}
\end{eqnarray}
i.e., the sites of the same sublattice-type do not interact
with each other, and
\begin{eqnarray}
(A^{-1})_{uv}^{\alpha\beta} & = & \langle \psi_{v}^{\beta}
\psi_u^{\alpha} \rangle =  
\left( \begin{array}{cc}
\displaystyle{a\over u-v} & 
\displaystyle{a\over u-{\bar v}} \\
\displaystyle{a\over {\bar u}-v} & 
\displaystyle{a\over {\bar u}-{\bar v}}
\end{array} \right)
\label{2.10c} \\
(A^{-1})_{vu}^{\alpha\beta} & = & \langle \psi_{u}^{\beta}
\psi_v^{\alpha} \rangle = 
\left( \begin{array}{cc}
\displaystyle{a\over v-u} & 
\displaystyle{a\over v-{\bar u}} \\
\displaystyle{a\over {\bar v}-u} & 
\displaystyle{a\over {\bar v}-{\bar u}}
\end{array} \right)
\label{2.10d}
\end{eqnarray}
\end{subequations}
with the required antisymmetry property $(A^{-1})_{uv}^{\alpha\beta}
= - (A^{-1})_{vu}^{\beta\alpha}$.
Here, the rows and the columns of the matrices are numbered as 
$\alpha,\beta=1,2$.

We now introduce an antisymmetric ``mass'' matrix,
\begin{subequations} \label{2.11}
\begin{eqnarray}
M_{uu'}^{\alpha\alpha'} & = & \delta_{uu'}
\left( \begin{array}{cc}
0 & {\rm i}\zeta(u) \\
-{\rm i} \zeta(u) & 0
\end{array} \right)
\label{2.11a} \\
M_{vv'}^{\alpha\alpha'} & = & \delta_{vv'}
\left( \begin{array}{cc}
0 & {\rm i}\zeta(v) \\
-{\rm i} \zeta(v) & 0
\end{array} \right)
\label{2.11b} \\
M_{uv}^{\alpha\beta} & = & 0 \label{2.11c} \\
M_{vu}^{\alpha\beta} & = & 0 \label{2.11d} 
\end{eqnarray}
\end{subequations}
Define
\begin{subequations} \label{2.12}
\begin{equation} \label{2.12a}
Z = \int {\rm d}\theta ~ {\rm e}^{-S}
\end{equation}
with the action
\begin{eqnarray} 
-S & = & {1\over 2}\ ^t\theta ({\bf A} + {\bf M}) \theta \nonumber \\
& = & - S_0 +\sum_u {\rm i} \zeta(u) \psi_u^1 \psi_u^2
+\sum_v {\rm i} \zeta(v) \psi_v^1 \psi_v^2  \label{2.12b}
\end{eqnarray}
\end{subequations}
Clearly,
\begin{equation} \label{2.13}
Z = {\rm Pf}({\bf A} + {\bf M})
\end{equation}
Let us expand the ratio $Z/Z_0$ in fugacities
$\{ \zeta(u) \}$ and $\{ \zeta(v) \}$. 
Owing to (\ref{2.10}), only neutral contributions with an
equal number of $\{ \zeta(u) \}$ and $\{ \zeta(v) \}$
will survive,
\begin{equation} \label{2.14}
{Z\over Z_0} = 1 + \sum_{N=1}^{\infty} (-1)^N 
\sum_{u_1<u_2<\ldots<u_N\in U \atop v_1<v_2<\ldots<v_N\in V}
\prod_{i=1}^N \zeta(u_i) \zeta(v_i) 
\left\langle \prod_{i=1}^N \left( \psi^1_{u_i} \psi^2_{u_i} 
\psi^1_{v_i} \psi^2_{v_i} \right) \right\rangle  
\end{equation}
With the definition of pair averages (\ref{2.10}),
the Wick theorem for fermions (see, e.g, \cite{Zinn-Justin},
\cite{Ramond}) implies
\begin{eqnarray} \label{2.15}
\left\langle \prod_{i=1}^N \left( \psi^1_{u_i} \psi^2_{u_i} 
\psi^1_{v_i} \psi^2_{v_i} \right) \right\rangle & = & (-1)^N
\left\langle \prod_{i=1}^N \left( \psi^1_{v_i} \psi^1_{u_i} 
\psi^2_{v_i} \psi^2_{u_i} \right) \right\rangle \nonumber \\
& = & (-1)^N   
{\rm det} 
\left( \begin{array}{cc}
\displaystyle{a\over u_i-v_j} & 
\displaystyle{a\over u_i-{\bar v}_j} \\
\displaystyle{a\over {\bar u}_i-v_j} & 
\displaystyle{a\over {\bar u}_i-{\bar v}_j}
\end{array} \right)_{i,j=1,\ldots,N}
\end{eqnarray} 
Comparing with (\ref{2.3}) and (\ref{2.4}) one concludes that
\begin{equation} \label{2.16}
\Xi = {Z\over Z_0} = {{\rm Pf}({\bf A} + {\bf M}) \over
{\rm Pf}({\bf A})}
\end{equation}
Introducing the matrix ${\bf K} = {\bf M}{\bf A}^{-1}$,
(\ref{2.16}) can be rewritten as
\begin{equation} \label{2.17}
\ln \Xi = {1\over 2} \ln {\rm det} (1 + {\bf K})
= {1\over 2} {\rm Tr} \ln (1 + {\bf K})
\end{equation}
Marking the charge sign of the particle at ${\vek r} = (z,{\bar z})$
by an index $s=\pm$, i.e., identifying $u=({\vek r},+)$,
$v=({\vek r},-)$ and $\zeta(u) = \zeta_+({\vek r})$,
$\zeta(v) = \zeta_-({\vek r})$, the ${\bf K}$-matrix has the elements
\begin{equation} \label{2.18}
K_{ss'}^{\alpha\alpha'}({\vek r},{\vek r}')
= \delta_{s,-s'} ~ {\rm i} \zeta_s({\vek r}) ~
\left( \begin{array}{cc}
\displaystyle{a\over {\bar z}-z'} & 
\displaystyle{a\over {\bar z}-{\bar z}'} \\
\displaystyle{-a\over z-z'} & 
\displaystyle{-a\over z-{\bar z}'}
\end{array} \right)
\end{equation}
where the $\alpha, \alpha'=1,2$ indices label the elements of the
$2\times 2$ matrix.

\subsection{Many-particle densities}
The calculation of many-body densities from the generator of type
(\ref{2.17}) is straightforward (see, e.g., \cite{Cornu1}).
When $S$ is the area of a lattice cell, using the equality
\begin{equation} \label{2.19}
\zeta_{s_1}({\vek r}_1) {\partial \over \partial 
\zeta_{s_1}({\vek r}_1)} K_{ss'}^{\alpha\alpha'}({\vek r},{\vek r}')
= \delta_{s_1,s} \delta_{{\vekexp r}_1,{\vekexp r}}
K_{ss'}^{\alpha\alpha'}({\vek r},{\vek r}')
\end{equation}
one finds for the density of particles of one sign (number of
such particles per unit area)
\begin{eqnarray} \label{2.20}
n_{s_1}({\vek r}_1) & = & {1\over S} \zeta_{s_1}({\vek r}_1)
{\partial \over \partial \zeta_{s_1}({\vek r}_1)}
\ln \Xi ~ \Bigg\vert_{\zeta_s({\vekexp r})=\zeta} \nonumber \\
& = & {1\over 2} m \sum_{\alpha_1} 
G_{s_1s_1}^{\alpha_1\alpha_1}({\vek r}_1,{\vek r}_1)
\end{eqnarray}
Here, the matrix
\begin{equation} \label{2.21}
{\bf G} = {1\over 2\pi a \zeta} {{\bf K} \over 1+{\bf K}}
\end{equation}
is defined with $K$-elements evaluated at constant fugacity,
$\zeta_s({\vek r}) = \zeta$, and the rescaled fugacity $m$
is defined by $m = 2\pi a \zeta/S$.
Using
\begin{equation} \label{2.22}
{\partial \over \partial \zeta_{s_2}({\vek r}_2)}
{1\over 1+{\bf K}} = - ~ {1\over 1+{\bf K}}~ {\partial {\bf K} 
\over \partial \zeta_{s_2}({\vek r}_2)} ~ {1\over 1+{\bf K}}
\end{equation}
one finds for the truncated two-body density
\begin{eqnarray} \label{2.23}
n_{s_1s_2}^{(2)T}({\vek r}_1,{\vek r}_2) & = & {1\over S^2} 
\zeta_{s_1}({\vek r}_1) \zeta_{s_2}({\vek r}_2)
{\partial^2 \over \partial \zeta_{s_1}({\vek r}_1)
\partial \zeta_{s_2}({\vek r}_2)}
\ln \Xi ~ \Bigg\vert_{\zeta_s({\vekexp r})=\zeta} \nonumber \\
& = & - {1\over 2} m^2 \sum_{\alpha_1,\alpha_2} 
G_{s_1s_2}^{\alpha_1\alpha_2}({\vek r}_1,{\vek r}_2)
G_{s_2s_1}^{\alpha_2\alpha_1}({\vek r}_2,{\vek r}_1)
\end{eqnarray}
By successive iterations, one finds for the truncated
$k$-body density
\begin{equation} \label{2.24}
n^{(k)T}_{s_1\ldots s_k}({\vek r}_1,\ldots,{\vek r}_k) = 
(-1)^{k+1}{1\over 2} m^k \sum_{\alpha_1,\ldots,\alpha_k}
\sum_{(i_1 i_2\ldots i_k)} 
G_{s_{i_1}s_{i_2}}^{\alpha_{i_1}\alpha_{i_2}}
({\vek r}_{i_1},{\vek r}_{i_2}) \ldots
G_{s_{i_k}s_{i_1}}^{\alpha_{i_k}\alpha_{i_1}}
({\vek r}_{i_k},{\vek r}_{i_1})
\end{equation}
where the second summation runs over all cycles 
$(i_1 i_2 \ldots i_k)$ built with $\{ 1,2,\ldots,k\}$.

In the continuum limit where the lattice spacing goes to zero,
the ${\bf G}_{ss'}$ $(s, s' = \pm)$ matrices, defined by
(\ref{2.21}), satisfy the integral equations
\begin{subequations} \label{2.25}
\begin{eqnarray}
{\bf G}_{ss}({\vek r}_1,{\vek r}_2) & = & - {\rm i}
\left( {m\over 2\pi} \right) \int_D {\rm d}^2 r
\left( \begin{array}{cc}
\displaystyle{1\over {\bar z}_1-z} & 
\displaystyle{1\over {\bar z}_1-{\bar z}} \\
\displaystyle{-1\over z_1-z} & 
\displaystyle{-1\over z_1-{\bar z}}
\end{array} \right)
\cdot {\bf G}_{-s,s}({\vek r},{\vek r}_2) \label{2.25a} \\
{\bf G}_{-s,s}({\vek r}_1,{\vek r}_2) & = & 
{\rm i} \left( {1\over 2\pi} \right) 
\left( \begin{array}{cc}
\displaystyle{1\over {\bar z}_1-z_2} & 
\displaystyle{1\over {\bar z}_1-{\bar z}_2} \\
\displaystyle{-1\over z_1-z_2} & 
\displaystyle{-1\over z_1-{\bar z}_2}
\end{array} \right) \nonumber \\
& & - {\rm i} \left( {m\over 2\pi} \right) \int_D {\rm d}^2 r
\left( \begin{array}{cc}
\displaystyle{1\over {\bar z}_1-z} & 
\displaystyle{1\over {\bar z}_1-{\bar z}} \\
\displaystyle{-1\over z_1-z} & 
\displaystyle{-1\over z_1-{\bar z}}
\end{array} \right)
\cdot {\bf G}_{ss}({\vek r},{\vek r}_2) \label{2.25b}
\end{eqnarray}
\end{subequations}
where the domain of integration $D$ is the half plane $y\ge 0$.
The operators 
\begin{equation} \label{2.26}
\partial_z = {1\over 2} \left(
{\partial \over \partial x} - {\rm i} {\partial \over \partial y}
\right) , \quad \quad \quad
\partial_{\bar z} = {1\over 2} \left( {\partial \over \partial x} + 
{\rm i} {\partial \over \partial y} \right)
\end{equation}
act as follows \cite{Zinn-Justin}
\begin{equation} \label{2.27}
\partial_z {1\over {\bar z} - {\bar z}'} = \pi 
\delta({\vek r}-{\vek r}'),
\quad \quad \quad
\partial_{\bar z} {1\over z - z'} = \pi \delta({\vek r}-{\vek r}')
\end{equation}
Differentiating appropriately equations (\ref{2.25}), one gets
\begin{subequations} \label{2.28}
\begin{eqnarray}
\left( \begin{array}{cc}
0 & \partial_{{\bar z}_1} \\
-\partial_{z_1} & 0
\end{array} \right)
{\bf G}_{ss}({\vek r}_1,{\vek r}_2) & = & {\rm i} \left( {m\over 2} 
\right) {\bf G}_{-s,s}({\vek r}_1,{\vek r}_2) \label{2.28a} \\
\left( \begin{array}{cc}
0 & \partial_{{\bar z}_1} \\
-\partial_{z_1} & 0
\end{array} \right)
{\bf G}_{-s,s}({\vek r}_1,{\vek r}_2) & = & - {\rm i}
{1\over 2} \delta({\vek r}_1 - {\vek r}_2) {\bf 1} + 
{\rm i} \left( {m\over 2} \right)
{\bf G}_{ss}({\vek r}_1,{\vek r}_2) 
\label{2.28b}
\end{eqnarray}
\end{subequations}
where ${\bf 1}$ denotes the $2\times 2$ unit matrix.
Using $\partial_z\partial_{{\bar z}} = (1/4) \Delta$, these
equations can be combined into
\begin{equation} \label{2.29}
( - \Delta_1 + m^2 ) {\bf G}_{ss}({\vek r}_1,{\vek r}_2)
= m \delta({\vek r}_1-{\vek r}_2) {\bf 1}
\end{equation}
The boundary conditions are given by the integral equations
(\ref{2.25}):
\begin{subequations} \label{2.30}
\begin{eqnarray}
G_{ss'}^{11}({\vek r}_1,{\vek r}_2) +
G_{ss'}^{21}({\vek r}_1,{\vek r}_2) & = & 0 , 
\quad \quad \quad ({\vek r}_1 \vee {\vek r}_2) \in \partial D
\label{2.30a} \\ 
G_{ss'}^{12}({\vek r}_1,{\vek r}_2) +
G_{ss'}^{22}({\vek r}_1,{\vek r}_2) & = & 0 ,
\quad \quad \quad ({\vek r}_1 \vee {\vek r}_2) \in \partial D
\label{2.30b}
\end{eqnarray}
\end{subequations}
where $\partial D$ is the domain boundary $y=0$.

Since ${\bf G}_{ss'}({\vek r}_1,{\vek r}_2)$ is translationally 
invariant along the $x$ axis, for solving the partial differential 
equations (\ref{2.28}) with the boundary conditions (\ref{2.30}), 
it is appropriate to introduce the Fourier transform 
with respect to $x_1-x_2$ defined by 
\begin{equation} \label{2.31}
{\bf G}_{ss'}({\vek r}_1,{\vek r}_2) = \int_{-\infty}^{\infty}
\frac{{\rm d}l}{2\pi} \tilde{\bf G}_{ss'}(y_1,y_2,l)
{\rm e}^{{\rm i}l(x_1-x_2)}
\end{equation}
We then obtain ordinary differential equations in $y_1$. 
Eq. (\ref{2.29}) becomes
\begin{equation} \label {2.32}
\left[-\frac{\partial^2}{\partial y_1^2}+\kappa^2\right]
\tilde{\bf G}_{ss}(y_1,y_2,l)=m\delta(y_1-y_2)\mathbf{1}
\end{equation}
where $\kappa =(m^2+l^2)^{1/2}$, and we look for a 
solution of (\ref{2.32}), symmetric in $y_1$ and $y_2$, in which a
``reflected'' part is added to the  free space solution, 
i.e. of the form 
\begin{equation} \label{2.33}
\tilde{G}_{ss}^{\alpha \alpha'}(y_1,y_2,l)=\frac{m}{2\kappa}
[\delta_{\alpha \alpha'}{\mathrm e}^{-\kappa |y_1-y_2|}
+A^{\alpha \alpha'}(l){\mathrm e}^{-\kappa (y_1+y_2)}]
\end{equation}
where $A^{\alpha \alpha'}(l)$ are functions to be determined. 
The matrix elements $G_{-s,s}^{\alpha \alpha'}$ are related to 
(\ref{2.33}) by the Fourier transforms of the relations (\ref{2.28a}), 
and the Fourier transforms of the boundary conditions (\ref{2.30}) 
determine the coefficients $A^{\alpha \alpha'}(l)$ in (\ref{2.33}). 
The result is
\begin{subequations}
\begin{eqnarray} 
\tilde{G}_{ss}^{11}(y_1,y_2,l) & = &\frac{m}{2\kappa}
\left[{\rm e}^{-\kappa |y_1-y_2|}+\left(\frac{\kappa}{l}-1\right)
{\rm e}^{-\kappa (y_1+y_2)} \right] \label{2.34a} \\
\tilde{G}_{ss}^{22}(y_1,y_2,l) & = &\frac{m}{2\kappa}  
\left[{\rm e}^{-\kappa |y_1-y_2|}- \left(\frac{\kappa}{l}+1\right)
{\rm e}^{-\kappa (y_1+y_2)} \right] \label{2.34b} \\
\tilde{G}_{ss}^{12}(y_1,y_2,l) & = &\frac{m}{2l}
{\rm e}^{-\kappa (y_1+y_2)} \label{2.34c} \\
\tilde{G}_{ss}^{21}(y_1,y_2,l) & = &-\frac{m}{2l}
{\rm e}^{-\kappa (y_1+y_2)} \label{2.34d}
\end{eqnarray}
\end{subequations}
The inverse Fourier transforms are
\begin{subequations} \label{2.35}
\begin{eqnarray}
G_{ss}^{11}({\vek r}_1,{\vek r}_2)&=&\frac{m}{2\pi}[K_0(mr_{12})
-K_0(mr_{12}^*)+{\mathrm i}I(x_1-x_2,y_1+y_2)] \\
G_{ss}^{22}({\vek r}_1,{\vek r}_2)&=&\frac{m}{2\pi}[K_0(mr_{12})
-K_0(mr_{12}^*)-{\mathrm i}I(x_1-x_2,y_1+y_2)] \\
G_{ss}^{12}({\vek r}_1,{\vek r}_2)&=&{\mathrm i}\frac{m}{2\pi}
I(x_1-x_2,y_1+y_2) \\
G_{ss}^{21}({\vek r}_1,{\vek r}_2)&=&-{\mathrm i}\frac{m}{2\pi}
I(x_1-x_2,y_1+y_2) 
\end{eqnarray}
\end{subequations}
where $r_{12}=|{\vek r}_1-{\vek r}_2|,\: r_{12}^*=|{\vek r}_1
-{\vek r}_2^*|$ 
with ${\vek r}_2^*=(x_2,-y_2)$ the image of ${\vek r}_2,\: K_0$ is a
modified Bessel function, and $I$ is the function (obtained as
a principal value)
\begin{equation} \label{2.36}
I(x_1-x_2,y_1+y_2)=\int_0^{\infty}{\rm d}l ~
\frac{\sin[l(x_1-x_2)]}{l} 
{\mathrm e}^{-\kappa (y_1+y_2)}
\end{equation}
Using the Fourier transforms of (\ref{2.28a}) we obtain the $(-s,s)$
matrix elements
\begin{subequations} \label{2.37}
\begin{eqnarray} 
\tilde{G}_{-s,s}^{11}(y_1,y_2,l) & = & \frac{1}{2\kappa}
\left( -\kappa +l+\frac{m^2}{l} \right)
{\rm e}^{-\kappa (y_1+y_2)} \\
\tilde{G}_{-s,s}^{22}(y_1,y_2,l) & = & \frac{1}{2\kappa}  
\left( -\kappa -l-\frac{m^2}{l} \right)
{\rm e}^{-\kappa (y_1+y_2)} \\
\tilde{G}_{-s,s}^{12}(y_1,y_2,l) & = & \frac{1}{2\kappa}
\left( \left[ l-\kappa ~ {\rm sign}(y_1-y_2) \right] 
{\rm e}^{-\kappa |y_1-y_2|}
+\frac{m^2}{l}{\rm e}^{-\kappa (y_1+y_2)} \right) \\
\tilde{G}_{-s,s}^{21}(y_1,y_2,l) & = & \frac{1}{2\kappa}
\left( \left[ -l-\kappa ~ {\rm sign}(y_1-y_2) \right] 
{\rm e}^{-\kappa |y_1-y_2|}
-\frac{m^2}{l}{\rm e}^{-\kappa (y_1+y_2)} \right)
\end{eqnarray}
\end{subequations}
The inverse Fourier transforms are
\begin{subequations} \label{2.38}
\begin{eqnarray}
G_{-s,s}^{11}({\vek r}_1,{\vek r}_2)&=&\frac{m}{2\pi}\left[
\frac{{\rm i}(x_1-x_2)-(y_1+y_2)}{r_{12}^*}K_1(mr_{12}^*)
+{\mathrm i}J(x_1-x_2,y_1+y_2)\right] \quad \quad \\
G_{-s,s}^{22}({\vek r}_1,{\vek r}_2)&=&\frac{m}{2\pi}\left[
\frac{-{\rm i}(x_1-x_2)-(y_1+y_2)}{r_{12}^*}K_1(mr_{12}^*)
-{\mathrm i}J(x_1-x_2,y_1+y_2)\right] \quad \quad \\
G_{-s,s}^{12}({\vek r}_1,{\vek r}_2)&=&\frac{m}{2\pi}\left[
\frac{{\rm i}(x_1-x_2)-(y_1-y_2)}{r_{12}}K_1(mr_{12})
+{\mathrm i}J(x_1-x_2,y_1+y_2)\right] \quad \quad \\
G_{-s,s}^{21}({\vek r}_1,{\vek r}_2)&=&\frac{m}{2\pi}\left[
\frac{-{\rm i}(x_1-x_2)-(y_1-y_2)}{r_{12}}K_1(mr_{12})
-{\rm i}J(x_1-x_2,y_1+y_2)\right] \quad \quad
\end{eqnarray}
\end{subequations}
where $K_1$ is a modified Bessel function and $J$ is the function
\begin{equation} \label{2.39}
J(x_1-x_2,y_1+y_2)=\int_0^{\infty}{\rm d}l ~
\frac{m\sin[l(x_1-x_2)]}{\kappa l}{\rm e}^{-\kappa (y_1+y_2)}
\end{equation}

Since $K_0(mr_{12})$ diverges logarithmically as $r_{12}\rightarrow 0$,
using (\ref{2.35}) in (\ref{2.20}) gives divergent densities, like in
the free space problem. 
However, if we subtract from $G_{ss}^{\alpha \alpha}$ 
its free space value, we get a finite result for the
difference $n_s(y)-n_s(\infty)$ (a density depends only on $y$):
\begin{equation} \label{2.40}  
n_s(y)-n_s(\infty)=-\frac{m^2}{2\pi}K_0(2my)
\end{equation}

Using (\ref{2.35}) and (\ref{2.38}) in (\ref{2.23}), we get the
truncated two-body densities
\begin{subequations} \label{2.41}
\begin{eqnarray}
n_{ss}^{(2)T}({\vek r}_1,{\vek r}_2)&=&-\frac{m^2}{2\pi}
[K_0(mr_{12})-K_0(mr_{12}^*)]^2 \\
n_{-s,s}^{(2)T}({\vek r}_1,{\vek r}_2)&=&\frac{m^2}{2\pi}
\Bigg\{ [K_1(mr_{12})]^2-[K_1(mr_{12}^*)]^2 \nonumber \\
& & + 2J(x_1-x_2,y_1+y_2)(x_1-x_2) \left[ \frac{K_1(mr_{12})}{r_{12}}
-\frac{K_1(mr_{12}^*)}{r_{12}^*} \right] \Bigg\} \quad
\end{eqnarray}
\end{subequations}

Writing (formally) the (infinite) bulk densities as $n_s(\infty)=
(m^2/2\pi)K_0(0)$, we see on (\ref{2.40}) that $n_s(0)=0$, as expected
because of the strong repulsion between a particle and its image.
Similarly, one checks on (\ref{2.41}) that also the two-body densities
vanish, as expected, when one of the particles is on the boundary.

As a consequence of the assumption of perfect screening, the charge
correlation functions are expected to obey a variety of sum rules. 
In particular, an even multipole moment of a particle plus its
surrounding screening cloud should vanish \cite{Martin} (in the presence
of an ideal dielectric boundary, the situation is different for odd
multipole moments: some of their components trivially vanish for
symmetry reasons, while other components have no a priori reason for
vanishing since they are cancelled by the images). 
As a check of our results (\ref{2.40}) and (\ref{2.41}) for the one 
and two-body densities, we explicitly derive the monopole and 
quadrupole sum rules. 
Since the one-body densities are divergent, we subtract the 
corresponding bulk moment for the system in infinite space,
without a boundary. 
Thus, we expect the monopole sum rule 
\begin{eqnarray} \label{2.42}
& & \int_{y_1>0} {\rm d}^2r_1
\left[ n_{ss}^{(2)T}({\vek r}_1,{\vek r}_2)
-n_{s,-s}^{(2)T}({\vek r}_1,{\vek r}_2) 
-n_{ss}^{(2)T}(r_{12};\infty) \right.
\nonumber \\  & & \left.
-n_{ss}^{(2)T}(r_{12}^*;\infty) 
+n_{s,-s}^{(2)T}(r_{12};\infty)
+n_{s,-s}^{(2)T}(r_{12}^*;\infty) \right] =-n_s(y_2)+n_s(\infty)
\end{eqnarray}
and the quadrupole sum rule
\begin{eqnarray} \label{2.43}
& &\int_{y_1>0} {\rm d}^2r_1 \left[ y_1^2-(x_1-x_2)^2 \right]
\left[ n_{ss}^{(2)T} ({\vek r}_1,{\vek r}_2)
-n_{s,-s}^{(2)T}({\vek r}_1,{\vek r}_2) 
-n_{ss}^{(2)T}(r_{12};\infty) \right.
\nonumber \\ & &  \left.
-n_{ss}^{(2)T}(r_{12}^*;\infty) 
+n_{s,-s}^{(2)T}(r_{12};\infty)
+n_{s,-s}^{(2)T}(r_{12}^*;\infty) \right] =-y_2^2[n_s(y_2)-n_s(\infty)]
\end{eqnarray}
where $n_{ss}^{(2)T}(r_{12};\infty)=-(m^2/2\pi)^2[K_0(mr_{12})]^2$ and 
$n_{s,-s}^{(2)T}(r_{12};\infty)=(m^2/2\pi)^2[K_1(mr_{12})]^2$ are the
bulk truncated two-body densities (the moments of the system without
a boundary have been replaced by integrals over the half-space
$y_1>0$ and the addition of $n_{ss'}^{(2)T}(r_{12}^*;\infty)$ terms). 
These sum rules are derived in the Appendix.

\renewcommand{\theequation}{3.\arabic{equation}}
\setcounter{equation}{0}

\section{Strip}
We now consider a strip geometry.
The Coulomb fluid, infinite in the $x$-direction,
is constrained by two ideal dielectric walls at
$y=0$ and $y=W$.
The above studied half-plane case corresponds to
the limit $W\to \infty$.

The Coulomb potential between the dielectric walls
is the solution of the Poisson equation
$\Delta v({\vek r},{\vek r}') = 
- 2 \pi \delta({\vek r}-{\vek r}')$
with the boundary conditions for the $y$-component
of the electric field $E_y(x,0) = 0$ and $E_y(x,W) = 0$.
By the method of images \cite{Jackson}, after subtracting
an irrelevant infinite constant, one gets for
the Coulomb potential
\begin{equation} \label{3.1}
v({\vek r},{\vek r}') = - \ln \left\vert
{\sinh k(z-z') \over k a} ~ {\sinh k({\bar z}-z') \over k a}
\right\vert
\end{equation}
where $k=\pi/(2W)$.
Each particle has a self-interaction
$-(1/2)\ln \vert \sinh k(z-{\bar z})/(ka) \vert$.
Writing
\begin{equation} \label{3.2}
\sinh k(z-z') = {1\over 2} {\rm e}^{-k(z+z')} ({\rm e}^{2kz}
- {\rm e}^{2kz'})
\end{equation}
and using $\exp(2ku)$ and $\exp(2kv)$ instead of $u$ and $v$,
one can proceed as in the previous section, with the final
result for the grand partition function, at $\beta =2$,
\begin{equation} \label{3.3}
\ln \Xi = {1\over 2} {\rm Tr} \ln (1+{\bf K})
\end{equation}
where
\begin{equation} \label{3.4}
K_{ss'}^{\alpha\alpha'}({\vek r},{\vek r}')
= \delta_{s,-s'} ~ {\rm i} \zeta ~
\left( \begin{array}{cc}
\displaystyle{ka \over \sinh k({\bar z}-z')} & 
\displaystyle{ka \over \sinh k({\bar z}-{\bar z}')} \\
\displaystyle{-ka \over \sinh k(z-z')} & 
\displaystyle{-ka \over \sinh k(z-{\bar z}')}
\end{array} \right)
\end{equation}
In the limit $W\to\infty$ $(k\to 0)$, (\ref{3.4}) reduces to
(\ref{2.18}) as it should be.

In the continuum limit, keeping the previous definition of
the rescaled fugacity $m=2\pi a \zeta/S$, the eigenfunctions
$\{ \psi_s^{\alpha}({\vek r}) \}$ of $m^{-1} {\bf K}$ and 
the corresponding eigenvalues $1/\lambda$ are defined by
\begin{equation} \label{3.5}
{1\over m} \int_D {{\rm d}^2 r' \over S} \sum_{s'=\pm}
\sum_{\alpha'=1,2} K_{ss'}^{\alpha\alpha'}({\vek r},{\vek r}')
\psi_{s'}^{\alpha'}({\vek r}') = {1\over \lambda} 
\psi_s^{\alpha}({\vek r})
\end{equation} 
where the domain of integration $D$ is now the strip.
Eq. (\ref{3.5}) corresponds to a set of coupled integral 
equations
\begin{subequations} \label{3.6}
\begin{eqnarray}
{{\rm i}\lambda \over 4W} \int_D {\rm d}^2 r' \left[
{1\over \sinh k({\bar z}-z')} \psi_{-s}^1({\vek r}') +
{1\over \sinh k({\bar z}-{\bar z}')} \psi_{-s}^2({\vek r}')
\right] & = & \psi_s^1({\vek r}) \label{3.6a} \\
{-{\rm i}\lambda \over 4W} \int_D {\rm d}^2 r' \left[
{1\over \sinh k(z-z')} \psi_{-s}^1({\vek r}') +
{1\over \sinh k(z-{\bar z}')} \psi_{-s}^2({\vek r}')
\right] & = & \psi_s^2({\vek r}) \label{3.6b}
\end{eqnarray}
\end{subequations}
for $s = \pm$.
In terms of $\lambda$,
\begin{equation} \label{3.7}
\ln \Xi = {1\over 2} \sum_{\lambda} \ln \left( 
1 + {m\over \lambda} \right)
\end{equation}
By using the equalities
\begin{equation} \label{3.8}
\partial_z {1\over 
\sinh k({\bar z}-{\bar z}')} =
\partial_{{\bar z}} {1\over \sinh k(z-z')}
= {\pi \over k} \delta({\vek r}-{\vek r}'), \quad \quad
{\vek r}, {\vek r}' \in D 
\end{equation}
the integral equations (\ref{3.6}) can be transformed into
the differential equations
\begin{subequations} \label{3.9}
\begin{eqnarray}
\partial_z \psi_s^1({\vek r}) & = &
{{\rm i} \lambda \over 2} \psi_{-s}^2({\vek r}) \label{3.9a} \\
\partial_{{\bar z}} \psi_s^2({\vek r}) & = &
- {{\rm i} \lambda \over 2} \psi_{-s}^1 ({\vek r}) \label{3.9b}
\end{eqnarray}
\end{subequations}
The combination of these equations results in the
Laplacian eigenvalue problem
\begin{equation} \label{3.10}
( - \Delta + \lambda^2 ) \psi_s^{\alpha}({\vek r}) = 0
\end{equation}
with boundary conditions given by the integral equations 
(\ref{3.6}):
\begin{equation} \label{3.11}
\psi_s^1(x,0) = - \psi_s^2(x,0) \quad \quad {\rm and}
\quad \quad \psi_s^1(x,W) = \psi_s^2(x,W)
\end{equation}

For the present geometry, we look for a solution which is
translationally invariant along the $x$-axis, i.e.,
\begin{equation} \label{3.12}
\psi_s^{\alpha}(x,y) = {\rm e}^{{\rm i}lx} \left( 
A_s^{\alpha} {\rm e}^{\kappa y} + 
B_s^{\alpha} {\rm e}^{-\kappa y} \right)
\end{equation}
where $\kappa = (l^2+m^2)^{1/2}$.
Due to the relations (\ref{3.9}), only four of eight
coefficients $\{ A_s^{\alpha}, B_s^{\alpha} \}$ are
independent, say $\{ A_+^{\alpha}, B_+^{\alpha} \}$.
The boundary conditions (\ref{3.11}) imply for them
a system of four linear homogeneous equations.
The existence of a nonvanishing solution is given
by the nullity of the determinant of this system.
This gives rise to the relation between $\lambda$ and $l$:
\begin{equation} \label{3.13}
\cosh \left[ W (l^2+\lambda^2)^{1/2} \right] \pm
\lambda {\sinh \left[ W (l^2+\lambda^2)^{1/2} \right]
\over (l^2+\lambda^2)^{1/2}} = 0
\end{equation}
The $\pm$ sign means that, for a given $l$, two
$\lambda$'s with opposite signs occur.
Let us define the entire function
\begin{equation} \label{3.14} 
f(z) = {1\over \cosh(Wl)} \left(
\cosh \left[ W (l^2+z^2)^{1/2} \right] -
z {\sinh \left[ W (l^2+z^2)^{1/2} \right]
\over (l^2+z^2)^{1/2}} \right)
\end{equation}
The solutions of (\ref{3.13}) are the zeros of
$f(\pm z)$.
Since $f(0) = 1$, we have
\begin{equation} \label{3.15}
f(z) = \prod_{\lambda \in f^{-1}(0)} \left( 
1 - {z\over \lambda} \right)
\end{equation}
Therefore, from (\ref{3.7}), the grand potential per
unit length $\omega$ is given by
\begin{eqnarray} \label{3.16}
\beta \omega & = & - {1\over 2} \int_{-\infty}^{\infty}
{{\rm d}l \over 2\pi} \ln \prod_{\lambda\in f^{-1}(0)}
\left[ \left( 1 + {m\over \lambda} \right)
\left( 1 - {m\over \lambda} \right) \right] \nonumber \\
& = & - {1\over 2\pi} \int_0^{\infty} {\rm d}l ~ 
\ln [ f(-m) f(m) ]
\end{eqnarray}
In the limit $W\to\infty$, it holds
\begin{eqnarray} \label{3.17}
\ln f(\pm m) & = & W \left[ (l^2+m^2)^{1/2} - \vert l \vert \right]
+ \ln \left( 1 \mp {m\over (l^2+m^2)^{1/2}} \right) \nonumber \\
& & - \ln \left( 1 + {\rm e}^{-2\vert l \vert W} \right)
+ O\left( {\rm e}^{-mW} \right)
\end{eqnarray}
Introducing a short-range repulsion cutoff $l_{\max} = 1/\sigma$
in order to avoid the divergence of the bulk term, one has finally,
at $\beta=2$,
\begin{equation} \label{3.18}
\beta \omega = - \beta P W + 2 \beta \gamma + {\pi \over 24 W}
+ O\left( {\rm e}^{-mW} \right)
\end{equation}
where
\begin{equation} \label{3.19}
\beta P = {m^2 \over 2\pi} \left[ \ln \left( {2\over m\sigma}
\right) + 1 \right]
\end{equation}
and
\begin{equation} \label{3.20}
\beta \gamma = {m\over 4}
\end{equation}
$P$ is the bulk pressure of an infinite system \cite{Cornu1},
with a modified cutoff procedure, and $\gamma$ is the surface tension.
The leading finite-size correction term has the universal
form $\pi/(24W)$.
The same universal term, except for a change of sign, is found
for the massless Gaussian field theory defined on the strip,
with various boundary conditions of the conformally invariant
type \cite{Cardy}.

We notice that the surface tension $\gamma$ can be computed
directly from the density profile obtained in the half-space,
formula (\ref{2.40}).
$\gamma$ is the boundary part per unit length of the grand
potential $\Omega$.
The total number of particles is given by 
$N = N_+ + N_- = - \beta \zeta \partial \Omega / \partial \zeta$.
The boundary part of this relation is
\begin{equation} \label{3.21}
- \beta m {\partial \over \partial m} \gamma =
\int_0^{\infty} {\rm d} y ~ [ n(y) - n]  
\end{equation}
where the total particle density $n(y) = n_+(y) + n_-(y)$
and $n = n_+(\infty) + n_-(\infty)$, 
and we have used that $m=2\pi \zeta$.
With respect to (\ref{2.40}), it holds 
\begin{equation} \label{3.22}
\int_0^{\infty} {\rm d} y ~ [ n(y) - n] = - {m\over 4}
\end{equation}
Inserting this into (\ref{3.21}), one rederives the formula (\ref{3.20}).
This result is also reproduced by the exact solution of the surface
tension \cite{Samaj1} (valid for an arbitrary $\beta<3$ and obtained 
by using completely different means), evaluated at $\beta = 2$.

\renewcommand{\theequation}{4.\arabic{equation}}
\setcounter{equation}{0}

\section{Universality of the finite-size correction}
The finite-size correction to the grand potential, eq. (\ref{3.18}) of the 
previous section, actually is a special case of a very general result
valid at any temperature and in any dimension $d\:(d\geq 2)$, for a
conducting Coulomb system confined between two parallel ideal dielectric
plates separated by a distance $W$; the Coulomb system extends to
infinity in the $d-1$ other directions. We shall show that the grand
potential $\omega$ per unit area of one plate (times the inverse
temperature $\beta$) has the large-$W$ expansion
\begin{equation} \label{4.1}
\beta\omega = -\beta P W +2\beta\gamma +\frac{C(d)}{W^{d-1}} + \ldots
\end{equation}
where $P$ is the bulk pressure and $\gamma$ the surface tension; these
quantities are non-universal. However, the last term of (\ref{4.1}) is a 
universal finite-size correction, with a coefficient $C(d)$ depending
only on the dimension $d$:
\begin{equation} \label{4.2}
C(d)=\frac{\Gamma (d/2)\,\zeta(d)}{2^d\pi^{d/2}} 
\end{equation}
where $\Gamma$ is the Gamma function and $\zeta$ the Riemann zeta
function. In particular, $C(2)=\pi /24$.

This universal finite-size correction is of the same nature as the ones
which occur when the electric potential obeys periodic boundary
conditions on the plates \cite{Forrester3} or when the plates are ideal 
conductors (Dirichlet boundary conditions) \cite{Jancovici}.  
It is remarkable that $C(d)$ has the same value for ideal conductor and
ideal dielectric plates.

In the present case of ideal dielectric plates, two different
derivations  of (\ref{4.1}) can be obtained by minor changes in the
derivations which have already been made in the case of ideal conductor
plates \cite{Jancovici}. 
Here, we shall concentrate on that derivation which relies on 
the assumption that the Coulomb system exhibits perfect
screening properties. 
Therefore, the universal finite-size correction is  
\emph{not} expected to hold in the absence of such screening properties,
for instance in the low-temperature Kosterlitz-Thouless phase of a
two-dimensional Coulomb gas. 
It should also be noted that if some short-range interactions 
are added to the Coulomb ones, the screening properties and 
therefore (\ref{4.1}) are still expected to hold.

For deriving (\ref{4.1}) from the screening properties of the Coulomb
system, closely following \cite{Jancovici}, 
we first establish a sum rule. 
The $d$-dimensional Coulomb system is supposed to fill the
slab $0<y<W$, with ideal dielectric plates at $y=0$ and $y=W$.  
Let $\hat{E}_x(0)$ be a Cartesian component parallel to the plates of
the microscopic electric field produced by the plasma at some point on 
the plate $y=0$, say the origin, and let $\hat{\rho} ({\vek r})$ 
be the microscopic charge density at some point in the Coulomb system. 
If an external infinitesimal dipole $p$, oriented parallel to the
plates, say along the $x$ axis, is placed at the origin (on the Coulomb 
system side), since the system is assumed to have good screening 
properties it responds through the appearance of an induced charge 
density $\delta \rho ({\vek r})$, localized near the origin and having 
a dipole moment opposite to $p$: 
\begin{equation} \label{4.3}
\int {\mathrm d}^2r\,x\,\delta \rho ({\vek r})=-p
\end{equation}
(choosing $p$ parallel to the plates insures that it has to be screened
by the Coulomb system itself, in spite of the presence of images). 
On the other hand, the interaction Hamiltonian between $p$ and the 
Coulomb system is $-p\hat{E}_x(0)$ and linear response theory gives
$\delta \rho({\vek r}) = \beta p \langle \hat{E}_x(0) 
\hat{\rho}({\vek r}) \rangle^T$, where $\langle \ldots \rangle^T$ 
denotes a truncated two-point function of the unperturbed system. 
Thus, the correlation function obeys the sum rule
\begin{equation} \label{4.4}
\beta \int {\rm d}^2r\, x \langle \hat{E}_x(0)\hat{\rho}({\vek r})
\rangle^T = -1
\end{equation}

We now compute the force per unit area acting on the plate $y=0$. Let
$\mu_d =\linebreak (d-2)2\pi^{d/2}/\Gamma (d/2)$ if $d>2$, 
$\mu_2=2\pi$. The unit of charge is defined such that the Coulomb
interaction between two unit charges in infinite space be 
$v_0({\vek r}-{\vek r}')=|{\vek r}-{\vek r}'|^{d-2}$ if
$d>2$ and $-\ln (|{\vek r}-{\vek r}'|/a)$ if $d=2$. Then, the Coulomb
potential between the ideal dielectric plates can be written as a sum
over images  
\begin{equation} \label{4.5}
v({\vek r},{\vek r}')=\sum_{n=-\infty}^{n=\infty}
[v_0({\vek r}+n2W{\vek u}-{\vek r}')+v_0({\vek r}^*+n2W{\vek u}-
{\vek r}')] 
\end{equation} 
where ${\vek u}$ is the unit vector of the $y$ axis and ${\vek r}^*
={\vek r}-2y{\vek u}$ is an image of {\vek r}. 
Actually, the sum in (\ref{4.5}) does not converge, but it can be 
made finite through the subtraction of some (infinite) constant 
which we do not write explicitly since it is irrelevant in what
follows. 
The force per unit area acting on the plate $y=0$ has only a component 
along the $y$ axis, which is the $yy$ component of the statistical
average of the microscopic Maxwell stress tensor  
\begin{equation} \label{4.6}
T_{yy}(0)=\frac{1}{\mu_d} \langle \hat{E}_y(0)^2-
\frac{1}{2}\hat{{\vek E}}({\vek r})^2 \rangle 
=-\frac{(d-1)}{2\mu_d} \langle \hat{E}_x(0)^2 \rangle^T
\end{equation}
where we have used that, on the ideal dielectric plate,
$\hat{E}_y(0)=0$ while all components of $\hat{{\vek E}}({\vek r})$
parallel to the plate give the same contribution; also, by charge
symmetry, the average electric field vanishes and therefore 
$\langle \ldots \rangle$ can be replaced by $\langle \ldots \rangle^T$. 
Since the density $n(y)$ vanishes on an ideal dielectric plate 
(because of the strong particle-image repulsion), the force on the 
plate has no contact contribution $n(0)kT$. Using
\begin{equation} \label{4.7}
\hat{E}_x(0)=-\int {\rm d}^2 r ~ \frac{\partial v({\vek r},{\vek r}')}
{\partial x'} \Big\vert_{{\vekexp r}'=0}\, \hat{\rho}({\vek r})
\end{equation}
(\ref{4.6}) can be rewritten as
\begin{equation} \label{4.8}
T_{yy}(0)=\frac{(d-1)}{2\mu_d}\int {\rm d}^2r ~
\frac{\partial v({\vek r},{\vek r}')}{\partial x'}
\Big\vert_{{\vekexp r}'=0} \langle \hat{E}_x(0)\hat{\rho}
({\vek r}) \rangle^T
\end{equation}

As the distance $W$ between the plates increases,  
$\partial v/\partial x'$ can be expanded in powers of $W^{-1}$. 
Using (\ref{4.5}) in (\ref{4.8}) we find
\begin{equation} \label{4.9}
T_{yy}(0)=T_{yy}(0)\vert_{W=\infty}+\frac{(d-1)\Gamma(d/2)\zeta (d)}
{2^d\pi^{d/2}W^d}\int {\rm d}^2r\, x \langle \hat{E}_x(0)\hat{\rho}
({\vek r}) \rangle^T+O \left( \frac{1}{W^{d+1}} \right)
\end{equation}
The integral in (\ref{4.9}) obeys the sum rule (\ref{4.4}). Since 
$\partial\omega/\partial W=T_{yy}(0)$, (\ref{4.9}) gives 
(\ref{4.1}) and (\ref{4.2}). 

In the peculiar case of a one-component plasma, due to the presence of a
neutralizing background, the bulk term in the grand potential $\omega$
is not of the form $-PW$. Also the force on a plate 
has an additional term related to the potential difference between the 
surface and the bulk of the plasma \cite{Choquard}, \cite{Totsuji}. 
However that additional term does not contribute to the universal
finite-size correction which keeps the same form 
\footnote{Although they have been inadvertently omitted in
ref.\cite{Jancovici}, the same remarks apply to the case of ideal
conductor plates.}. 

\section{Concluding remarks}
The model under consideration was a two-component Coulomb gas
in contact with walls made of ideal dielectric material.
As shown in section 2, the model is solvable, in two
dimensions at inverse temperature $\beta =2$, by using the
Pfaffian method.
This means that, not the grand partition function, but its 
square is expressible as a ``treatable'' determinant.
This is the fundamental difference with the previously
solved cases of the Coulomb gas in contact with a plain hard wall
\cite{Cornu2}, or an ideal conductor wall 
\cite{Cornu2} -- \cite{Jancovici},
or with periodic boundary conditions on the plates \cite{Forrester3}.
As a consequence, the introduction of a four-component (instead
of two-component) Fermi field, associated with each point of space,
is necessary.
We have worked with matrices which elements are themselves
$2\times 2$ matrices, so without saying it we have used a
variant of the algebra of quaternion matrices.

For the rectilinear geometry of a semi-infinite dielectric wall,
we have computed the particle densities (\ref{2.40}) and 
the correlation functions (\ref{2.41}).
Due to the screening effect, the correlations
are supposed to obey a variety of sum rules \cite{Martin}.
In the Appendix, we have checked the monopole (\ref{2.42}) and
quadrupole (\ref{2.43}) sum rules, with subtraction of
(divergent) bulk moments.
The relatively complicated form of the truncated pair
correlations prevents us, in practice, to go beyond the
verification of these sum rules.

The strip formalism in section 3 is technically very similar 
to the one for ideal-conductor boundaries \cite{Jancovici}.
The main formal difference is that, when calculating the
grand potential per unit length (\ref{3.16}), an ``average''
$(1/2)[\ln f(-m) + \ln f(m)]$ instead of $\ln f(-m)$ 
should be integrated.
This makes the surface tension finite, see formula (\ref{3.20}).
The universal finite-size correction term can also be obtained,
as a consequence of the good screening properties, in the more
general case of a Coulomb system of arbitrary dimension
$d$ $(d\ge 2)$ confined in a slab of width $W$, at an arbitrary
temperature.
The universal finite-size correction has the same value for
ideal-conductor and ideal-dielectric walls.
In two dimensions, the $\pi/(24W)$ correction term also
appears in some papers about the sine-Gordon theory
\cite{Skorik}; thus, one might suspect that the integrability
of these theories is not a necessary ingredient for obtaining
this universal term, and a universal term might also be present
in sine-Gordon models of higher dimension although they are
not integrable. 

\renewcommand{\theequation}{A.\arabic{equation}}
\setcounter{equation}{0}

\section*{Appendix}

In this Appendix we sketch the derivation of the sum rules (\ref{2.42}) 
and (\ref{2.43}). Without loss of generality, we can choose $x_2=0$.

The l.h.s. of (\ref{2.42}) is
\begin{eqnarray} \label{A.1}
F_1 & = & \left( \frac{m^2}{2\pi} \right)^2 \int_0^{\infty}
{\rm d}y_1 \int_{-\infty}^{\infty}{\rm d}x_1
\Bigg\{ K_0(mr_{12})\,K_0(mr_{12}^*)
+[K_1(mr_{12}^*)]^2 \nonumber \\
&&-2J(x_1,y_1+y_2)x_1 \left[ \frac{K_1(mr_{12})}{r_{12}}-
\frac{K_1(mr_{12}^*)}{r_{12}^*} \right] \Bigg\}
\end{eqnarray}
where $J(x_1,y_1+y_2)$, as defined by (\ref{2.39}), has the
properties 
\begin{equation} \label{A.2}
J(0,y_1+y_2)=0
\end{equation} 
and
\begin{equation} \label{A.3} 
\frac{\partial J}{\partial x_1}=\int_0^{\infty}{\rm d}l
~ \frac{m\cos(lx_1)}{\kappa}{\mathrm e}^{-\kappa (y_1+y_2)}
=mK_0(mr_{12})
\end{equation} 
Since
\begin{equation} \label{A.4}
m x_1 \left[ \frac{K_1(mr_{12})}{r_{12}}-\frac{K_1(mr_{12}^*)}{r_{12}^*}
\right] = - \frac{\partial}{\partial x_1}[K_0(mr_{12})-K_0(mr_{12}^*)] 
\end{equation}
an integration per partes on $x_1$ transforms (\ref{A.1}) into
\begin{equation} \label{A.5} 
F_1 = 2 \left( \frac{m^2}{2\pi} \right)^2 \int_0^{\infty}{\rm d}y_1
\int_{-\infty}^{\infty}{\rm d} x_1 \left\{ [K_0(mr_{12}^*)]^2
+[K_1(mr_{12}^*)]^2 \right\}
\end{equation}
When $K_0$ and $K_1$ in (\ref{A.5}) are replaced by their Fourier
transforms with respect to $x_1$
\begin{subequations} \label{A.6}
\begin{equation}
K_0(mr_{12}^*) = \int_{-\infty}^{\infty}{\rm d}l ~ {\rm e}^{{\rm i}lx_1}
\frac{1}{2\kappa}{\mathrm e}^{-\kappa (y_1+y_2)}
\end{equation}
and
\begin{equation}
\frac{\pm{\mathrm i}x_1+y_1-y_2}{r_{12}}K_1(mr_{12}^*)=
\int_{-\infty}^{\infty}{\mathrm d}l ~ {\rm e}^{\pm{\mathrm i}lx_1}
\frac{l+\kappa}{2m\kappa} {\rm e}^{-\kappa (y_1+y_2)}
\end{equation}
\end{subequations}
(\ref{A.5}) becomes
\begin{equation} \label{A.7}
F_1=\frac{m^2}{2\pi}\int_0^{\infty}{\mathrm d}y_1\int_{-\infty}^{\infty}
{\mathrm d}l ~ {\mathrm e}^{-2\kappa (y_1+y_2)}
\end{equation} 
Performing first the integration on $y_1$, we obtain
\begin{equation} \label{A.8}
F_1=\frac{m^2}{2\pi}\int_{-\infty}^{\infty}{\mathrm d}l ~
\frac{1}{2\kappa}{\mathrm e}^{-2\kappa y_2}=\frac{m^2}{2\pi}K_0(2my_2) 
\end{equation}
With regard to (\ref{2.40}), (\ref{A.8}) gives the monopole sum rule
(\ref{2.42}).

The l.h.s. of (\ref{2.43}) is
\begin{eqnarray} \label{A.9}
F_2 & = & \left( \frac{m^2}{2\pi} \right)^2
\int_0^{\infty}{\rm d} y_1
\int_{-\infty}^{\infty}{\rm d} x_1 (y_1^2-x_1^2)
\Bigg\{ K_0(mr_{12})\,K_0(mr_{12}^*)
+[K_1(mr_{12}^*]^2 \nonumber \\
&&-2J(x_1,y_1+y_2)x_1 \left[
\frac{K_1(mr_{12})}{r_{12}}-\frac{K_1(mr_{12}^*)}{r_{12}^*} \right]
\Bigg\}
\end{eqnarray}
Now, after an integration per partes on $x_1$, a $J$-dependent term is
left:
\begin{eqnarray} \label{A.10}
F_2 & = & 2 \left( \frac{m^2}{2\pi} \right)^2
\int_0^{\infty}{\rm d}y_1 
\int_{-\infty}^{\infty}{\mathrm d}x_1 \Bigg\{ (y_1^2-x_1^2)
\left([K_0(mr_{12}^*)]^2+[K_1(mr_{12}^*)]^2\right)\nonumber\\
&&+\frac{2}{m}J(x_1,y_1+y_2)\,x_1
\left[ K_0(mr_{12})-K_0(mr_{12}^*) \right] \Bigg\}
\end{eqnarray}
Again, by the introduction of appropriate Fourier transforms with
respect to $x_1$ (now we also need the Fourier transform of 
$x_1K_0(mr_{12})$, etc \ldots), the integral on $x_1$ is replaced by an
integral on $l$, and the integration on $y_1$ can be performed
first. After a straightforward but tedious calculation, the detail of
which we omit, one obtains
\begin{equation} \label{A.11}
F_2=\frac{m^2}{2\pi}\int_0^{\infty}{\rm d}l
\left[\frac{m^2-l^2}{\kappa^4}y_2+\frac{m^2-l^2}{\kappa^3}y_2^2\right]
{\rm e}^{-2\kappa y_2}
\end{equation}
An integration per partes transforms the term $(m^2-l^2)y_2/\kappa^4$ 
into $2l^2y_2^2/\kappa^3$. Thus
\begin{equation} \label{A.12}
F_2=y_2^2\frac{m^2}{2\pi}\int_0^{\infty}\frac{{\rm d}l}{\kappa} ~
{\rm e}^{-2\kappa y_2}=y_2^2\frac{m^2}{2\pi}K_0(2my_2)
\end{equation}
With regard to (\ref{2.40}), (\ref{A.12}) gives the quadrupole sum rule 
(\ref{2.43}).

\section*{Acknowledgments}
We are grateful to Fr{\'e}d{\'e}ric van Wijland for very
useful discussions.
The stay of L. {\v S}. to LPT Orsay 
is supported by a NATO fellowship.
A partial support by Grant VEGA 2/7174/20 is acknowledged.

\end{document}